\documentclass[aps,prl,preprint]{revtex4-1}
\usepackage{graphicx}
\usepackage{dcolumn}
\usepackage{bm}
\begin{document}
\title{Measuring the mode volume of plasmonic nanocavities using coupled optical emitters}
\author{Kasey J. Russell}
\author{Kitty Y. M. Yeung}
\author{Evelyn Hu}
\affiliation{Harvard University School of Engineering and Applied Sciences, Cambridge, MA 02138, USA}
\begin{abstract}
Metallic optical systems can confine light to deep sub-wavelength dimensions, but verifying the level of confinement at these length scales typically requires specialized techniques and equipment for probing the near-field of the structure. We experimentally measured the confinement of a metal-based optical cavity by using the cavity modes themselves as a sensitive probe of the cavity characteristics. By perturbing the cavity modes with conformal dielectric layers of sub-nm thickness using atomic layer deposition, we find the exponential decay length of the modes to be less than 5\% of the free-space wavelength ($\lambda$) and the mode volume to be of order $\lambda^3$/1000. These results provide experimental confirmation of the deep sub-wavelength confinement capabilities of metal-based optical cavities.
\end{abstract}
\maketitle

\section{I. Introduction}
A central goal of cavity quantum electrodynamics (cQED) is to alter the density of optical states--in both space and frequency--at the location of an emitter\cite{vahala_optical_2003,noda_spontaneous-emission_2007}. The motivation is especially clear in the weak-coupling regime, also known as the Purcell regime, in which Fermi's golden rule can be used to show that the spontaneous emission rate of an emitter is directly proportional to the local density of optical states. To date, a particular focus of the field has been developing dielectric structures that can achieve high degrees of spectral confinement, as characterized by the cavity quality (Q)\cite{vahala_optical_2003,noda_spontaneous-emission_2007}. At the other extreme, drawing on advances in Raman scattering and the rising interest in plasmonics, it has recently been shown that metal-optical structures should be able to achieve sufficient spatial confinement to achieve cQED effects such as enhanced spontaneous emission despite the poor spectral confinement (i.e. low Q) that results from using lossy metals at optical frequencies\cite{chang_quantum_2006,maier_effective_2006,gong_design_2007,jun_nonresonant_2008,esteban_optical_2010,vesseur_broadband_2010,miyazaki_squeezing_2006,gong_plasmonic_2009,jun_strong_2010,russell_large_2012}.

Cavity Q can be measured using a variety of techniques, including directly from a far-field emission spectrum. In contrast, measuring the spatial confinement typically requires specialized techniques such as scanning near-field optical microscopy (SNOM)\cite{cubukcu_plasmonic_2006,spasenovic_measurements_2009}; photon localization microscopy\cite{mcleod_nonperturbative_2011}; cathodoluminescence (CL)\cite{coenen_directional_2011}; or coating cavities in photo-sensitive polymers\cite{sundaramurthy_toward_2006,nah_metal-enhanced_2010}. Several of these techniques can yield detailed maps of relative field intensity, but each suffers from drawbacks such as requiring expensive equipment (as in the case of SNOM, CL, and photon localization microscopy) or being difficult to quantify (photo-polymerization).

Here, we present a more general and more sensitive approach to measuring the confinement of metal-based optical cavities. By coating the cavity in successive conformal layers of dielectric, we red-shift the cavity resonances by an amount proportional to the strength of the field at that layer. Ultimately, this approach yields a measurement of the evanescent decay of the cavity modes in structures with high aspect ratio and occluding top structures. The confinement can be measured with nanometer resolution, although the measured values of confinement are not strictly equal to those of the original cavity because the modes are perturbed by the deposited dielectric layer. This approach could be viewed as a variant of the approach used by Zhang et al.~\cite{zhang_far-field_2010}, in which the dependence of plasmonic resonances on cavity geometry was used to determine confinement. An important distinction between that work and the one presented here is that our approach can be used to measure confinement of individual cavities. In addition, this approach may prove useful for tuning the cavity resonances with respect to the emission lines of optical emitters for cQED investigations.

\section{II. Experimental}
The fabrication of the cavities used in this investigation has been described previously\cite{russell_gap-mode_2010}. The key criterion governing this cavity design was control over the cavity characteristics. This concern required that the metal surfaces be smooth (less than 1 nm rms) and that the critical dimension of the cavity--the gap between metals--be uniform. A silver substrate of sub-nm rms roughness was prepared via template stripping from an atomically flat Si wafer. This substrate was covered in a dielectric stack consisting of 5 nm sputtered SiN, one to two monolayers of PbS colloidal quantum dots (Evident), and 5 nm sputtered SiN. Silver nanowires (Blue Nano) of diameter $\sim$100 nm and length $\sim$1-30 $\mu$m were then deposited in a droplet of ethanol that was allowed to dry. A schematic diagram of a cavity is shown in Fig. 1.
\begin{figure}[htp]
\includegraphics{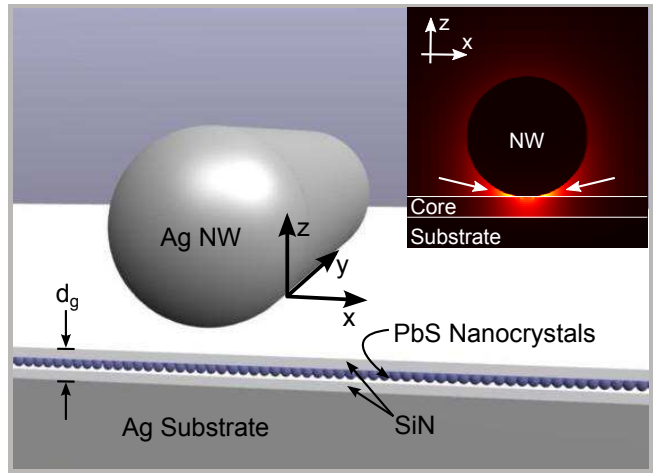}
\caption{Schematic diagram of cavity with dielectric core comprised of SiN/PbS nanocrystals/SiN. Typical nanowire (NW) length and diameter are $\sim$1 $\mu$m and 100 nm, respectively. Typical gap spacing $d_g$ is 15-20 nm. Inset, Cross-sectional plot in the x-z plane of simulated electric field profile in a cavity with 20 nm gap. Part of the cavity mode is confined to the area adjacent to the line of contact between the NW and dielectric core, as indicated with arrows.}
\end{figure}
\begin{figure*}[htp]
\includegraphics[scale=0.9]{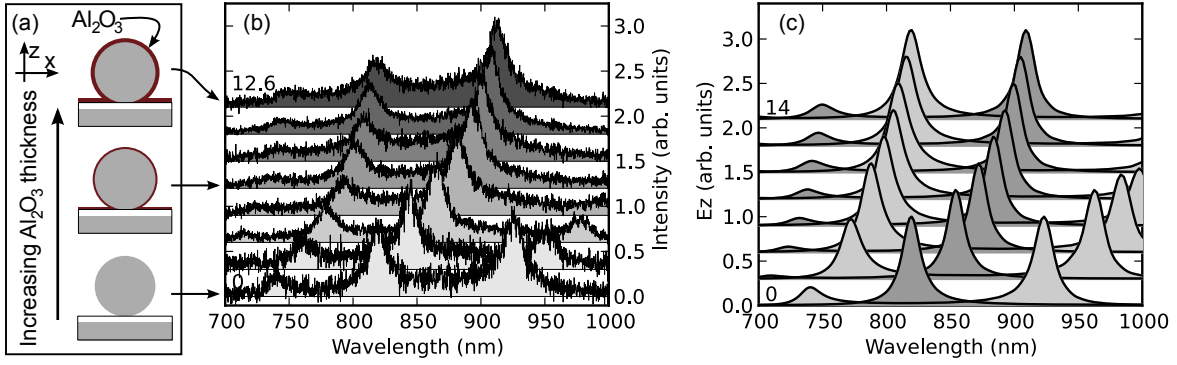}
\caption{(a), Schematic illustration of conformal atomic layer deposition of Al$_2$O$_3$ on a nanowire cavity. (b), Photoluminescence measurements of a single cavity after successive $\sim$1.8 nm depositions of Al$_2$O$_3$, vertically offset for clarity. Al$_2$O$_3$ thickness in nm is indicated for first and last measurements. Nanowire length and diameter were 1.28 $\pm$ 0.01 $\mu$m and 105 $\pm$ 10 nm, respectively. (c), FDTD simulated resonance spectra of a cavity with the same nanowire geometry as in (b). Different traces correspond to different thicknesses of conformal dielectric coating, vertically offset for clarity. Coating thickness increases in 2 nm increments from 0 nm to 14 nm (indicated, in nm, for first and last traces). }
\end{figure*}

To probe cavity confinement, emission spectra of individual cavities were measured as conformal layers of Al$_2$O$_3$ were deposited in $\sim$1.8 nm increments using atomic layer deposition (ALD), as illustrated schematically in Fig. 2(a). ALD is a chemical deposition process in which sequential pulses of precursors (in our case, deionized water and trimethylaluminum) react on the substrate to form a single molecular layer (Al$_2$O$_3$, in our case)\cite{puurunen_surface_2005}. The self-limiting nature of this deposition method has been shown to yield very smooth and conformal coatings\cite{puurunen_surface_2005}. Deposition was performed in a commercial system (Savannah S200, Cambridge NanoTech) at 50 $^\circ$C, yielding individual layers of Al$_2$O$_3$ that were $\sim$0.12 nm thick with an index of n $\sim$ 1.5\cite{groner_low-temperature_2004}, as confirmed by a 115 cycle deposition on a freshly-cleaned Si wafer. Each $\sim$ 1.8 nm deposition consisted of 15 ALD cycles. Cavity emission spectra were measured using room temperature micro-photoluminescence. Light from a 532 nm continuous-wave laser was directed through a 100$\times$, 0.5 NA microscope objective onto the sample in the direction normal to the substrate (i.e. along the z-axis). Fluorescence from the optically-excited PbS quantum dots ($\sim$ 850 nm center wavelength, $\sim$ 200 nm spectral width) was collected through the same microscope objective and analyzed using a grating spectrometer.

Each deposition of Al$_2$O$_3$ led to a red-shift of the cavity modes (see, e.g., Fig. 2(b)). This red-shift occurred because the deposited Al$_2$O$_3$ displaced a small, thin volume of air and thereby increased the index refraction within that volume from 1.0 to $\sim$1.5. The effect of this index change on the frequencies of the cavity resonances was approximately proportional to the amplitude of the cavity mode within the deposited layer (assuming that the thickness of the deposited layer is sufficiently thin to cause negligible change in the modal profile). Although the majority of the field energy of the mode is confined within the metal layers and the SiN/PbS/SiN dielectric core of the cavity, a significant fraction of the mode extends into the region adjacent to the line of contact between the nanowire and the dielectric core (inset, Fig. 1), and it is this portion of the cavity modes that is perturbed by the deposited Al$_2$O$_3$. The magnitude of the red-shift therefore decreased exponentially for each successive layer as a result of the exponential decay of the cavity modes away from the cavity. Measuring the red-shift of the cavity resonances as a function of Al$_2$O$_3$ thickness therefore provided a direct measurement of the cavity confinement.

Measurements were performed on a total of 7 cavities, all of which showed similar resonance shifts. In addition, three-dimensional finite-difference time-domain (FDTD) simulations (Lumerical Solutions, Inc.) were in excellent agreement with the experimental measurements (see Fig. 2(c)). For these simulations, silver material parameters were taken from Johnson and Christy\cite{johnson_optical_1972}, the SiN/PbS/SiN layer was modeled as a uniform dielectric layer of index 1.6 and thickness 20 nm, and the Al$_2$O$_3$ coating layer was modeled as a fully conformal dielectric of index 1.5. The length and diameter of the modeled nanowire were 1.28 $\mu$m and 100 nm, respectively, matching the measured dimensions of the cavity in Fig 2(b).

\section{III. Discussion}
The peak wavelengths of all modes from all experimentally measured cavities were plotted versus ALD thickness, and the dependence was fit with a function describing exponential saturation:
\begin{equation}\label{eq_saturation}
f(d) = \lambda_0 + \Delta \lambda (1-e^{-d/d_0}),
\end{equation} where $d$ is the Al$_2$O$_3$ thickness, $d_0$ is the characteristic thickness of Al$_2$O$_3$ that describes the exponential saturation of the resonance shift, $\lambda_0$ is the wavelength of the un-shifted mode, and $\Delta \lambda$ is the saturation value of the shift, i.e. the shift that would be induced by infinite Al$_2$O$_3$ thickness. This simple formula accurately describes the dependence of the data on the Al$_2$O$_3$ thickness (Fig. 3).
\begin{figure}[htp]
\includegraphics{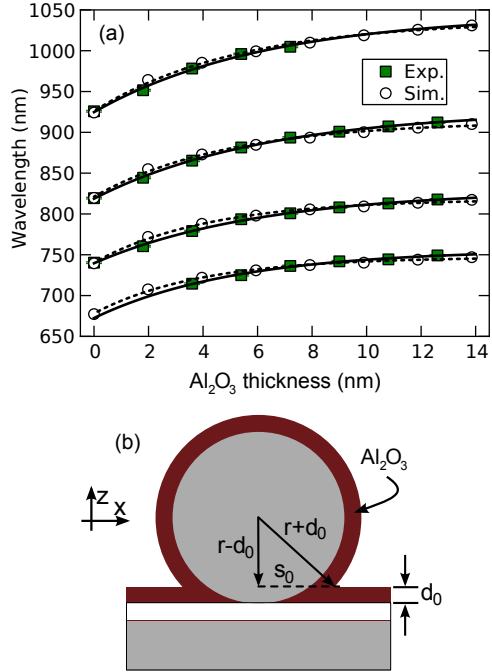}
\caption{(Color online) Example analysis of confinement from a single cavity. (a), Peak wavelengths extracted from Fig. 2(b) and (c) (Exp. and Sim., respectively) plotted versus thickness of ALD-deposited Al$_2$O$_3$. Solid and dashed lines are fits to Exp. and Sim., respectively, using Eq.~\ref{eq_saturation}. (b), Schematic model used to find the exponential decay length of the cavity mode, $s_0 = 2(rd_0)^{1/2}$, where $d_0$ is the characteristic ALD thickness found from the fit using Eq.~\ref{eq_saturation} and $r$ is the radius of the nanowire.}
\end{figure}

The characteristic thickness $d_0$ determined from fits using Eq.~\ref{eq_saturation} cannot simply be interpreted as a decay length of the cavity mode. This is for two reasons, both relating to the particular geometry of our cavity. First, there is no single characteristic decay length of the modes in our cavities: they decay evanescently in all directions away from the cavity, and the rate of decay depends strongly on position and direction relative to the metal surfaces. In particular, in the region of the cavity adjacent to the line of contact between the nanowire and dielectric core, the cavity modes decay rapidly away from the metal surfaces (i.e. in directions approximately parallel to the z-axis near the line of contact), and they decay more slowly in directions approximately parallel to the x-axis, such as directions parallel to the arrows in the inset of Fig. 1. For the purposes of measuring confinement, we are interested in the longest characteristic decay length of the cavity modes, and so this is what will be meant by the symbol $s_0$. Second, the thickness of Al$_2$O$_3$ is measured in a direction perpendicular to the surface on which it is deposited (i.e. in directions approximately parallel to the z-axis near the line of contact), and in our case this direction is not parallel to the direction in which we wish to measure the decay of the cavity modes (i.e. directions approximately parallel to the x-axis). As can be seen from the schematic diagram in Fig. 3(b), the lateral extent, $s$, of the Al$_2$O$_3$ can be approximated as $s \sim 2(rd)^{1/2}$, where $r$ is the radius of the nanowire (typically $r \sim$ 100 nm). This allows us to find the longest characteristic decay length of the cavity mode as $s_0 \sim 2(rd_0)^{1/2}$. Values for the fits shown in Fig. 3(a) are given in Table I.
\begin{table}[htp]
\begin{ruledtabular}
    \begin{tabular}{| l | l | l|}
	\hline
	$\lambda_0$ (nm) & $s_0$ (nm) & $\Delta \lambda/\lambda_0$ \\ \hline
	672.5 $\pm$ 9.9 & 33.3 $\pm$ 7.1 & 0.125 $\pm$ 0.010 \\
	739.6 $\pm$ 1.1 & 36.3 $\pm$ 2.4 & 0.121 $\pm$ 0.003 \\
	818.7 $\pm$ 0.8 & 36.7 $\pm$ 1.5 & 0.133 $\pm$ 0.002 \\
	924.8 $\pm$ 3.0 & 36.6 $\pm$ 10.8 & 0.130 $\pm$ 0.022
    \end{tabular}
\end{ruledtabular}
\caption{Fitting parameters for the fits shown in Fig. 3(a).}
\end{table}

	Fits were made to the modes of all measured cavities (24 modes from 7 cavities), yielding best-fit values for $\lambda_0$, $\Delta \lambda$ and $s_0$ for each mode (Fig. 4). Fig. 4(a) shows the distribution of characteristic decay lengths, which all lie in the range 30 nm $ < s_0 < $40 nm. These values of $s_0$ are only $\sim$ 5$\%$ of the free-space wavelength (the shortest-wavelength mode observed was $\sim$ 675 nm), yielding direct experimental evidence of the deep sub-wavelength confinement capabilities of our structure. Comparison can also be drawn between $s_0$ and the evanescent decay length into the dielectric of a surface plasmon polariton (SPP) at a single silver/dielectric interface, which is comparable to $\lambda$ at the frequencies considered here.
\begin{figure*}[htp]
\includegraphics{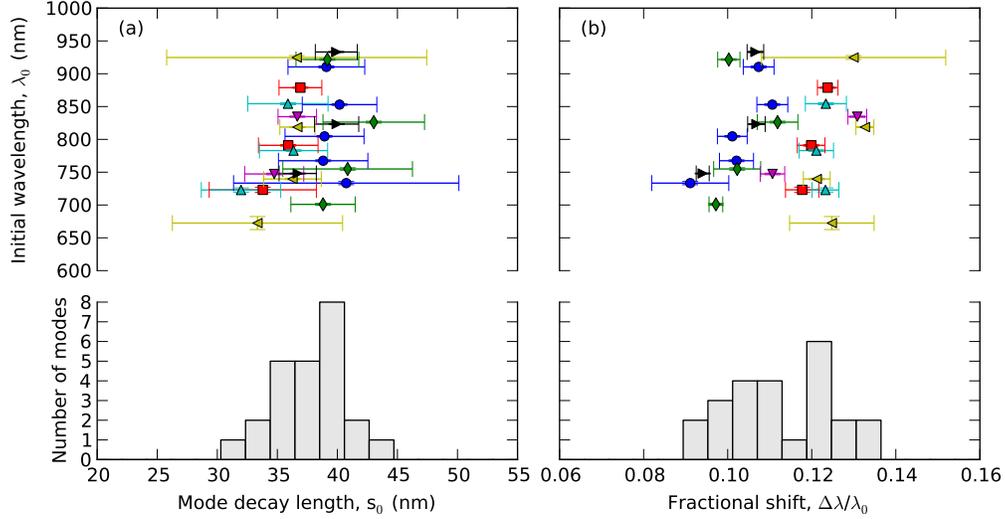}
\caption{(Color online) Experimental confinement characteristics extracted from fits to 24 modes from 7 different cavities using Eq.~\ref{eq_saturation} and Fig. 3(b). Modes from the same cavity are labeled with the same color and symbol. (a), Histogram and wavelength dependence of mode decay lengths, $s_0$. The average decay length is less than 5$\%$ of both the free-space wavelength and the corresponding decay length of a surface plasmon polariton at a single metal-dielectric interface. (b), Fractional shift $\Delta \lambda/\lambda_0$. The small fractional tuning confirms that the majority of the mode energy is confined within the metals and the SiN/PbS/SiN dielectric gap of the as-fabricated cavity.}
\end{figure*}

Shown in Fig. 4(b) are extracted values of fractional tuning, $\Delta \lambda/\lambda_0$. This quantity is monotonically related to the fraction of the mode energy contained in the deposited Al$_2$O$_3$ in the limit of infinite Al$_2$O$_3$ thickness. Quantitatively, it depends on the cavity geometry and how the mode profile is modified by the deposition of Al$_2$O$_3$. Qualitatively, a larger fractional tuning is observed when a greater fraction of the mode energy is contained in the Al$_2$O$_3$. For example, a resonator for surface plasmon polaritons (SPPs) at a single metal/dielectric interface would have $\Delta \lambda/\lambda_0$ $\sim$ 0.5 in the near-infrared spectral range. This is because more than 90$\%$ of the energy of an SPP at a single surface resides in the dielectric region\cite{maier_plasmonic_2006}.

In the case of our gap-mode cavity, we observe $\Delta \lambda/\lambda_0$ $\sim$ 0.12. The fractional tuning in our cavities is significantly smaller than in structures with a single metal/dielectric interface because the gap-like character of the modes in our cavities confines the majority of the mode within the metal layers and the SiN/PbS/SiN dielectric core (see Fig. 1 inset).

These two parameters--the measured evanescent decay length of the modes and the fractional shift of the cavity resonances--provide direct experimental evidence of the deep sub-wavelength confinement capabilities of plasmonic nanocavities. For the cavity measured in Fig 2(b), the measured decay length allows us to estimate the cavity mode volume as $V \sim 1.28 \mu$m $\times$ 0.08 $\mu$m $\times$ 0.02 $\mu$m = 2 $\times 10^{-3} \mu$m$^3 \sim 3 \times 10^{-3} \lambda^3$, where we have estimated the extent of the mode in the x-direction to be $\sim 2s_0$ and the extent in the z-direction to be approximately equal to the gap spacing $d_g \sim 20$ nm. The mode volume can also be normalized to the value of the wavelength within the gap dielectric, $\lambda_D = \lambda/n$, rather than to the free-space wavelength. This yields $V \sim 3 \times 10^{-3} (n\lambda_D)^3 \sim 1.2 \times 10^{-2} \lambda_D^3$, where we have used the estimate of $n \sim 1.6$ for the effective index of the SiN/PbS/SiN multilayer.

This experimental estimate of $V$ compares well with values calculated from FDTD simulations. For these calculations, we used the definition of Foresi \textit{et al.}\cite{foresi_photonic-bandgap_1997}, with the same minor correction used by Miyazaki and Kurokawa\cite{miyazaki_squeezing_2006}: the energy density was normalized not to its maximum value in the whole region but to the maximum within the center of the dielectric gap (the approximate location of the PbS nanocrystal layer). Discretized calculations such as FDTD can have large singular electric fields at the surfaces of metal regions that, when used to normalize the mode energy, result in unrealistically small values for $V$. With this prescription, we find $V \sim 1.3 \times 10^{-3} \mu$m$^3$ for the cavity simulated in Fig. 2(c) (without Al$_2$O$_3$ coating), in approximate agreement with the experimentally-determined estimates.

The ability to continuously vary the resonances of our nanocavity may have additional uses beyond metrology. When characterizing weak and strong cavity-emitter coupling, it is useful to spectrally tune the frequency of the cavity mode relative to that of the coupled emitter (see, e.g., Ref.~\cite{hennessy_quantum_2007}), and we are currently pursuing cavity designs that incorporate narrow-band emitters for this purpose.

\section{IV. Conclusion}
We have utilized the fluorescence of coupled broad-band optical emitters to characterize the confinement of a gap-mode plasmonic nanocavity. Our results show experimentally that our cavity confines light to deep sub-wavelength dimensions ($V \sim 3 \times 10^{-3} \lambda^3$), in agreement with numerical simulations. These results confirm that plasmonic structures are capable of strong confinement on the scale predicted by classical electrodynamics, paving the way for observation of large cQED effects in metal-optical devices.

We gratefully acknowledge helpful conversations with Tsung-Li Liu, financial support from NSF/NSEC under NSF/PHY-06-46094, the use of NSF/NNIN facilities at Harvard University’s Center for Nanoscale Systems, and the use of the hpc computer cluster at Harvard.

\end{document}